\documentclass[conference]{IEEEtran}
\usepackage{cite}

\ifCLASSINFOpdf
\else
\fi
\usepackage[dvips]{graphicx}
\usepackage{amsmath}
\usepackage{amssymb}
\usepackage{psfrag}

\newcommand{\defeq}{\stackrel{\rm def}{=}}
\newcommand{\vect}[1]{\boldsymbol{#1}}
\newtheorem{algo}{Algorithm}

\begin{document}
%
\title{A Steganography Based on \\CT-CDMA Communication Scheme \\
Using Complete Complementary Codes}

\author{\IEEEauthorblockN{Tetsuya Kojima}
\IEEEauthorblockA{Department of Computer Science\\
Tokyo National College of Technology\\
Hachioji, Tokyo, 193--0997 Japan\\
Email: kojt@ieee.org}
\and
\IEEEauthorblockN{Yoshiya Horii}
\IEEEauthorblockA{Advanced Course of Mechanical and\\
Computer Systems Engineering\\
Tokyo National College of Technology\\
Hachioji, Tokyo, 193--0997 Japan}}


%


\maketitle

\begin{abstract}
It has been shown that complete complementary codes can be 
applied into some communication systems like 
approximately synchronized CDMA systems 
because of its good correlation properties. 
CT-CDMA is one of the communication systems based on complete complementary 
codes. In this system, the information data of the multiple users 
can be transmitted by using the same set of complementary codes through 
a single frequency band. 
In this paper, we propose to apply CT-CDMA systems 
into a kind of steganography. 
It is shown that a large amount of secret 
data can be embedded in the stego image by the proposed method 
through some numerical experiments using color images. 
\end{abstract}


%
\IEEEpeerreviewmaketitle

\section{Introduction}

Various types of multimedia digital contents including images, music and 
videos have been distributed these days. 
In such circumstances, illegal copies of digital files have become 
big problems in today's network societies. 
On the other hand, there have been more and more demands to send 
the secret messages such as personal information securely. 
To solve these problems, the information hiding technologies 
including digital watermarks, steganography have become quite important 
and many research results have been reported. 
Recently, some researchers have reported 
the digital watermarking based on spread spectrum techniques 
\cite{cox97,hayashi2007,aminaga2003}. 

Complete complementary code is a kind of complementary codes 
proposed by Suehiro {\it et al.}\cite{suehiro88}, 
and has ideal auto- and cross-correlation properties. 
It is shown that complete complementary codes are effective to 
design the transmitted signal for some communication systems 
such as the approximately synchronized 
CDMA systems\cite{suehiro00,kojt2006,kojt2008}. 
Especially, in CT-CDMA 
(convoluted-time and code division multiple access) 
systems\cite{kojt2008}, 
different user's information data is overlapped and transmitted. 
In this systems, the information data of the multiple users can be 
transmitted by using a single auto-complementary codes 
through a single frequency band. 
In addition, if multiple auto-complementary codes are used 
at the same time, the large amount of the information data can be 
transmitted simultaneously through a single frequency band. 

The authors have proposed a method to apply complete complementary codes 
into the correlation-based digital watermarking 
for image data\cite{horii2009}. 
In this paper, as extension of the propsed digital watermarking scheme, 
we propose to apply CT-CDMA systems into a kind of steganography. 
In addition, it is shown that a large amount of secret 
data can be embedded in the cover image securely 
through numerical experiments.

\section{Preliminaries}

\subsection{Correlation Functions}

For any complex-valued sequence of the finite length $L$: 
\begin{equation}
  \vect{c_{a}} \defeq \{c_{a,0}, c_{a,1}, \ldots, c_{a,L-1}\}, 
\end{equation}
define the infinite length sequence $C_{a} \defeq \{ C_{a}(t) \}$ as 
\begin{equation}
  C_{a}(t) \defeq \sum_{i=0}^{L-1} c_{a,i} \delta_{i,t},  
\end{equation}
where $\delta_{i,n}$ denotes the Kronecker's delta. 
Note that the value of $C_{a}(t)$ is 0 for $t \leq -1, L \leq t$, 
so that $C_{a}(t)$ can be identified with $\vect{c_{a}}$. 
The cross-correlation function 
for any two finite-length sequences $\vect{c_{1}} =$ $\{c_{1,0},$ $c_{1,1},$ 
$\ldots,$ $c_{1,L-1}\}$ and $\vect{c_{2}} =$ $\{c_{2,0},$ $c_{2,1},$ $\ldots,$ 
$c_{2,L-1}\}$ are defined as 
\begin{equation}
  R_{C_{1},C_{2}}(\tau) 
	\defeq \sum_{t = -\infty}^{+ \infty} C_{1}(t) C_{2}^{*}(t-\tau),
\end{equation}
where $C_{a}^{*}(t)$ is the complex conjugate of $C_{a}(t)$. 
When $\vect{c_{1}} = \vect{c_{2}}$, 
the correlation function $R_{C_{1},C_{1}}(\tau)$ is 
called the auto-correlation function. 

\subsection{Complete Complementary Codes}

Complete complementary code is proposed 
by Suehiro {\it et al.}\cite{suehiro88}. 
It consists of several auto- complementary codes, 
any two of which are cross- complementary codes. 

For example, 
let the binary sequences $\vect{c_{0}^{(0)}}$, $\vect{c_{1}^{(0)}}$, 
$\vect{c_{0}^{(1)}}$, $\vect{c_{1}^{(1)}}$ 
be 
\begin{equation}
 \begin{array}{l l}
 	\vect{c_{0}^{(0)}} = \{- + - -\}, & \vect{c_{1}^{(0)}} = \{- - - +\},\\
 	\vect{c_{0}^{(1)}} = \{+ - - -\}, & \vect{c_{1}^{(1)}} = \{+ + - +\},
 \end{array}
 \label{eq:ex_ccc}
\end{equation}
where $+, -$ denotes $+1, -1$ respectively. 
The sum of the auto-correlation functions of $\vect{c_{0}^{(0)}}$ 
and $\vect{c_{1}^{(0)}}$ is 
\begin{equation}
  R_{C_{0}^{(0)}C_{0}^{(0)}}(\tau) + R_{C_{1}^{(0)}C_{1}^{(0)}}(u)
	= \left\{\begin{array}{rl}
		8, & {\rm if~} \tau = 0\\
		0, & {\rm if~} \tau \neq 0
		\end{array}\right..
\end{equation}
The similar property is also satisfied 
for $\vect{c_{0}^{(1)}}$ and $\vect{c_{1}^{(1)}}$. 
A set of sequences is called an auto-complementary code 
when the sum of their auto-correlation functions is 0 
in every term except for the zero shift term. 
Therefore, both of $\left\{\vect{c_{0}^{(0)}}, \vect{c_{1}^{(0)}}\right\}$ 
and $\left\{\vect{c_{0}^{(1)}}, \vect{c_{1}^{(1)}}\right\}$ are 
auto-complementary codes. 

On the other hand, the sums of the cross-correlation functions 
for these sequences can be written as 
\begin{equation}
  R_{C_{0}^{(0)}C_{0}^{(1)}}(\tau) + R_{C_{1}^{(0)}C_{1}^{(1)}}(\tau) = 0, 
 \label{eq:cross1}
\end{equation}
for any shift $-(L-1) \leq \tau \leq L-1$. 
If such properties are satisfied, 
a pair of the sequence sets $\left\{\vect{c_{0}^{(0)}},\right.$ 
$\left.\vect{c_{1}^{(0)}}\right\}$ 
and $\left\{\vect{c_{0}^{(1)}},\right.$ 
$\left.\vect{c_{1}^{(1)}} \right\}$ are called cross- complementary codes, 
and as a result, this couple of sets of finite length sequences 
$\left\{\left\{\vect{c_{0}^{(0)}}, \vect{c_{1}^{(0)}}\right\}, 
\left\{\vect{c_{0}^{(1)}}, \vect{c_{1}^{(1)}}\right\}\right\}$ is called 
a set of complete complementary codes. 

There are many variations among complete complementary codes\cite{jin2009}. 
In general, for the sequence length $l$, 
the number of sequences in each auto- complementary code $n$, 
and the number of different auto- complementary code $m$, 
the complete complementary code can be called 
as $(m, n, l)$- complete complementary code. 
The above example is a $(2,2,4)$- complete complementary code. 
An $(m, n, l)$- complete complementary code can be written as 
\begin{equation}
  \left\{\begin{array}{c c c c}
	\left\{ \vect{c_{0}^{(0)}},\right. & \vect{c_{1}^{(0)}}, 
		& \cdots, & \left. \vect{c_{n-1}^{(0)}} \right\} \\
	\left\{ \vect{c_{0}^{(1)}},\right. & \vect{c_{1}^{(1)}}, 
		& \cdots, & \left. \vect{c_{n-1}^{(1)}} \right\}\\
	\vdots & \vdots & & \vdots \\
	\left\{ \vect{c_{0}^{(m-1)}},\right. & \vect{c_{1}^{(m-1)}}, 
		& \cdots, & \left. \vect{c_{n-1}^{(m-1)}} \right\}
	\end{array}\right. , 
 \label{eq:mccc}
\end{equation}
where $m$ rows represent $m$ auto-complementary codes, 
any two of which are cross-complementary codes. 
The sum of the correlation functions satisfy the property 
such that, for any $i,k = 0, 1, \ldots, m-1$, 
\begin{equation}
  \sum_{j=0}^{n-1} R_{C_{j}^{(i)}C_{j}^{(k)}}(\tau)
	= \left\{\begin{array}{rl}
		A, & {\rm if~} i=k {\rm ~and~} \tau=0\\
		0, & {\rm otherwise}
		\end{array}\right., 
 \label{eq:correlation}
\end{equation}
where $A$ denotes a constant independent of the indices $i, k$.

\section{CT-CDMA}

\begin{figure}[!t]
  \begin{center}
	\includegraphics[scale=0.6]{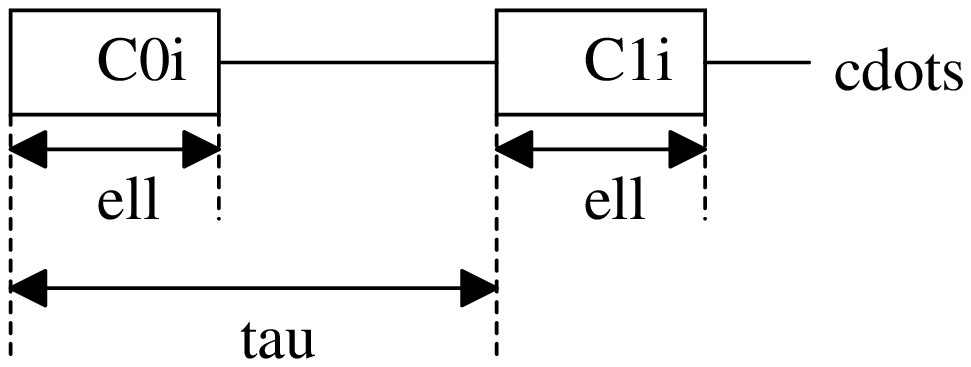}
  \end{center}
 \caption{Basic sequence.}
 \label{fig:basic}
\end{figure}

For any integer $i$ ($0 \leq i \leq m-1$), let 
\begin{equation}
  S^{(i)} \defeq \left\{s_{0}^{(i)}, s_{1}^{(i)}, \ldots, 
  	s_{n\tau -1}^{(i)}\right\} 
  	= \sum_{j=0}^{n-1} C_{j}^{(i)} \cdot Z^{-j \tau}, 
\end{equation}
be a basic sequence of the length $T = n \tau$, 
where $Z^{-k}$ denotes the delay operator of the length $k$, such that, 
$s_{t} \cdot Z^{-k} = s_{t-k}$ holds for any sequences $s_{t}$. 
Figure \ref{fig:basic} illustrates the basic sequence. 
\psfrag{C0i}{$\vect{c_{0}^{(i)}}$}
\psfrag{C1i}{$\vect{c_{1}^{(i)}}$}
\psfrag{ell}{$l$}
\psfrag{tau}{$\tau$}
\psfrag{cdots}{$\cdots$}
We consider the periodic sequence $\tilde{S}^{(i)} \defeq 
\left\{ \tilde{s}_{t}^{(i)} \right\}$ 
using the basic sequence $S^{(i)}$, such that, for any integer $k$, 
\begin{equation}
  \tilde{s}_{t + kT}^{(i)} = s_{t}^{(i)},
 \label{eq:periodic}
\end{equation}
where $T = n \tau$ is the period. 

Let $a \cdot \tilde{S}^{(0)}$ be a pilot signal for estimating 
the multipath channel property, where $a$ is a constant. 
On the other hand, the data symbol is modulated 
by the conventional DS-SS (direct sequence spread spectrum) method 
using the periodic sequence $\tilde{S}^{(1)}$. 
For example, if the $i$-th user transmits the $N$-bit data 
$\left( x_{0}^{(i)}, x_{1}^{(i)}, \ldots, x_{N-1}^{(i)} \right)$ 
$\in \{-1, +1\}^{N}$ and the information data signal $b_{t}^{(i)}$ 
for this user is given as 
\begin{equation}
  b_{t+k\tau}^{(i)} = x_{k}^{(i)}, 
 \label{eq:data_i}
\end{equation}
for any $k = 0, 1, \cdots, N-1$, 
the data signal is spread 
like $b_{t}^{(i)} \cdot \tilde{s}_{t}^{(1)}$. 

\begin{figure}[!t]
  \begin{center}
	\includegraphics[scale=0.56]{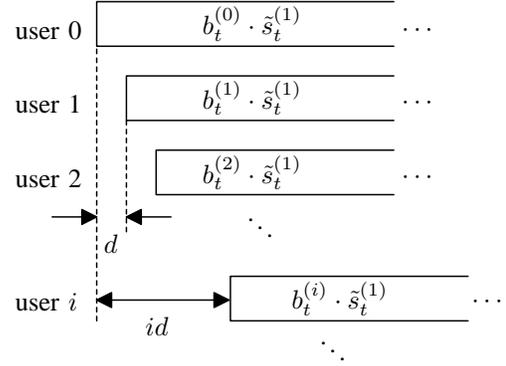}
  \end{center}
 \caption{Convoluted multi-user signals.}
 \label{fig:convolution}
\end{figure}

When $M$ users transmit the information data at the same time 
within a cell, each user's spread data signal is transmitted 
with the interval of $d$ [chip] in such a way as 
depicted in Fig. \ref{fig:convolution}. 
In other words, it can be considered that $M$ users' transmitted signal 
is convoluted as 
\begin{equation}
  \sum_{i=0}^{M-1} b_{t}^{(i)} \cdot \tilde{s}_{t}^{(1)} 
  		\cdot Z^{- i d}. 
 \label{eq:txsig}
\end{equation}
\psfrag{u0}{user 0}
\psfrag{u1}{user 1}
\psfrag{u2}{user 2}
\psfrag{ui}{user $i$}
\psfrag{signal0}{$b_{t}^{(0)} \cdot \tilde{s}_{t}^{(1)}$}
\psfrag{signal1}{$b_{t}^{(1)} \cdot \tilde{s}_{t}^{(1)}$}
\psfrag{signal2}{$b_{t}^{(2)} \cdot \tilde{s}_{t}^{(1)}$}
\psfrag{signali}{$b_{t}^{(i)} \cdot \tilde{s}_{t}^{(1)}$}
\psfrag{$d$}{$d$}
\psfrag{$id$}{$id$}
\psfrag{ddots}{$\ddots$}
All users' transmitters modulate their own spread data signals 
together with the common pilot signal 
by the same carrier of frequency and transmit them. 
The transmitted signal is received through the multipath channel. 

At the receiver, the received signal is input 
into the matched filter bank, where the sum of correlation functions between 
the received signals and the employed auto-correlation codes are evaluated. 
Observing the output of the matched filter bank, 
one can easily obtain the transmitted information data. 
See the reference \cite{kojt2008} for more details. 

In the above, only a pair of auto-complementary codes is used. 
However, we can use $(m/2)$ pairs of auto- complementary codes 
simultaneously as long as $(m, n, l)$-complete complementary codes 
are taken into account. 
Even if we assign another pair of auto-complementary codes 
for other users' transmissions, co-channel interference 
never occurs because of the superior 
characteristics of complete complementary codes given in 
Eq.(\ref{eq:correlation}). 
Therefore, the channels can be divided by 
both the convolution of the transmitted signals and the 
cross-correlation property of the complete complementary codes. 
CT-CDMA is named after this property \cite{kojt2008}. 

\section{Steganography Based on CT-CDMA}

Steganography is one of the information hiding technologies 
that transmitting the secret messages 
by embedding them into the given multimedia digital files such as 
music, still images, or videos. 
In this study, we propose a steganography based on CT-CDMA systems. 

\subsection{Basic Idea}

In CT-CDMA systems\cite{kojt2008}, 
the information data of the multiple users can be transmitted 
by using a single auto-complementary codes through a single frequency band. 
In addition, if multiple auto-complementary codes are used 
at the same time, the large amount of the information data can be 
transmitted simultaneously through a single frequency band. 

In this study, we apply this idea into a steganography. 
In order to embed a large amount of information, 
we propose to use the multiple auto-complementary codes 
included in a same complete complementary code. 
Each auto-complementary code can be considered as 
a mutually distinct communication channel. 
For steganography based on $(m,n,l)$-complete complementary codes, 
it is possible to use $m$ secret channels simultaneously at most. 
If $M$ ($\leq m$) channels are employed to embed the given secret message, 
the secret message is divided into $M$ fragments 
and embedded in $M$ distinct auto-complementary codes, that is, 
$M$ distinct secret channels. 

\subsection{Embedding Procedure}

In this paper, we consider that UTF-8 format text data 
of the size $MN$ bytes is embedded as the secret information 
into the cover image. A 24-bit bitmap color image is assumed 
as the cover image. 
The embedding procedure consists of the following 8 steps. 
\begin{algo}[Embedding Procedure]
\hspace*{1cm}\rm
\begin{enumerate}
  \item Transform the cover image from RGB format into 
	YCbCr format. 
  \item Transform each coefficient Y by two dimensional DCT.
  \item Extract the DCT components $\{d(t)\}$ 
	of the length $T = n\tau$ 
	$(0 \leq t \leq T-1)$ by using a secret key 
	from the middle frequency band of the DCT coefficients 
	evaluated by Step 2. 
  \item The secret message file is divided 
	into $M$ fragments of the size $N$ bytes. 
	Also, Each 1-byte character in the text file is divided into 
	the upper 4-bit sequence and the lower 4-bit sequence. 
	The obtained sequence of the length $2N$ is represented 
	by $X^{(i)} = \{$ $x_{0}^{(i)},$ $x_{1}^{(i)},$ $\ldots,$ 
	$x_{2N-1}^{(i)}\}$, $0 \leq i \leq M-1$. 
	Note that each $x_{t}^{(i)}$ takes the value 
	on $\{0, 1, \ldots, 15\}$. 
  \item The sequence $X^{(i)}$ is modulated by CT-CDMA. 
	The modulated CT-CDMA signal is represented 
	by $Y^{(i)} = \{ y_{t}^{(i)}\}$. 
  \item For $0 \leq \forall t \leq T-1$, 
	\begin{equation}
	  d'(t) = \alpha \sum_{i=0}^{M} y_{t}^{(i)} + d(t), 
	\end{equation}	
	where $\alpha$ $(> 0)$ is a given embedding coefficients 
	which controls the power of the embedded secret messages. 
  \item Put $d'(t)$ back into the DCT coefficients 
	instead of $d(t)$ for $0 \leq \forall t \leq T-1$. 
  \item Transform YCbCr format image into RGB format image. \\
	\hfill $\square$
\end{enumerate}
\end{algo}

\subsection{Extracting Procedure}

In the proposed method, the secret message is extracted 
from the stego image through the following 6 steps. 

\begin{algo}[Extracting Procedure]
\hspace*{1cm}\rm
\begin{enumerate}
  \item Transform the original and the stego images 
	from RGB format into YCbCr format. 
  \item Transform each coefficient Y of the original 
	and the stego images by two dimensional DCT. 
  \item Extract the DCT components $\{d^{o}(t)\}$ and \{$d^{s}(t)\}$ 
	of the length $T$ $(0 \leq t \leq T-1)$ from the original 
	and the stego images, respectively, 
	at the same frequency band used in the embedding procedure. 
  \item The difference sequence $D^{d} = \{d^{d}(t)\}$ 
	of the length $T$ is obtained in the following manner: 
	\begin{equation}
	  d^{d}(t) = d^{s}(t) - d^{o}(t), 
	\end{equation}
	where $0 \leq \forall t \leq T-1$. 
  \item Input the evaluated $d^{d}(t)$ into 
	the matched filter bank of the CT-CDMA system, 
	and obtain the sequence of the length $2N$, 
	$\widehat{X}^{(i)} = \{$ $\widehat{x}_{0}^{(i)},$ 
	$\widehat{x}_{1}^{(i)},$ $\ldots,$ 
	$\widehat{x}_{2N-1}^{(i)}\}$, $0 \leq i \leq M-1$. 
  \item From each two successive symbols 
	of the sequence $\widehat{X}^{(i)}$, 
	recover each 1-byte character of the embedded message. 
\end{enumerate}
\end{algo}

\subsection{Numerical Results}

In the following, 
$(16, 16, 256)$-complete complementary codes are employed to embed 
the secret message and the embedding coefficient $\alpha$ is 
given as $\alpha = 0.25$. We also set $T = 4096$ and $d = 1$ 
in CT-CDMA modulation. 
Image data is given as 24-bit bitmap file 
of the size $512 \times 512$ pixels. 

\begin{figure}[!t]
  \begin{center}
    \includegraphics[scale=0.65]{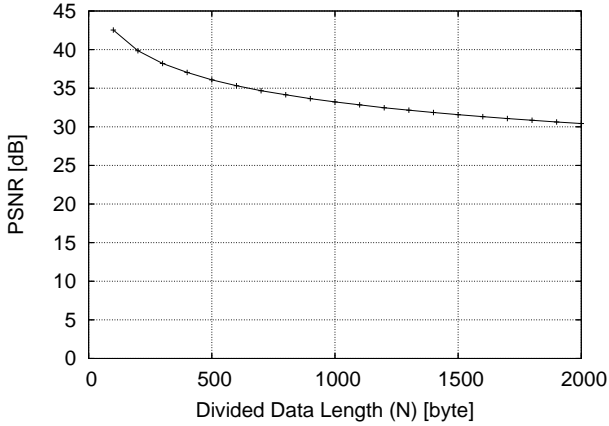}
  \end{center}
 \caption{The relation between the divided data length $N$ 
	and $PSNR$.}
 \label{fig:psnr1}
\end{figure}

Figure \ref{fig:psnr1} shows the relation between the length of 
the divided data fragment $N$ and $PSNR$. 
The number of secret channels is fixed as $M = 1$, 
and the divided data length is varied from $N = 100$ to 2,000 bytes. 
Each curve is evaluated by 100 independent trials. 
In general, it is known that it is quite difficult for human to 
recognize the difference between the cover and stego image 
if $PSNR$ is larger than around 35 [dB]. 
From Fig.\ref{fig:psnr1}, it can be seen that 
$PSNR \fallingdotseq 35$ [dB] when $N = 600$ [byte], 
so a message of the size 600 byte at most can be embedded 
secretly by using the proposed method. 

\begin{figure}[tb]
  \begin{center}
    \includegraphics[width=7.5cm,height=7.5cm]{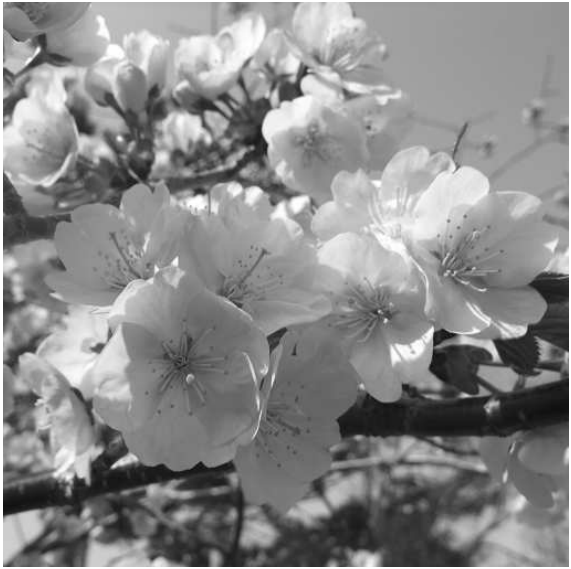}
  \end{center}
 \caption{Original image.}
 \label{fig:source}
\end{figure}

\begin{figure}[tb]
  \begin{center}
    \includegraphics[width=7.5cm,height=7.5cm]{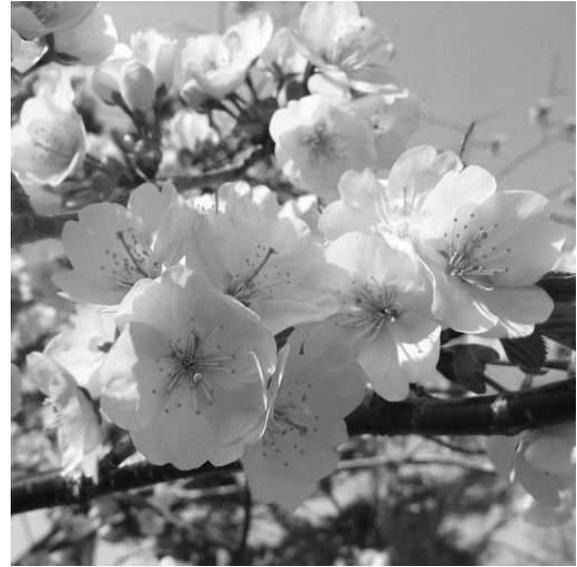}
  \end{center}
 \caption{Stego image $(N = 500, M = 1)$.}
 \label{fig:stego}
\end{figure}

The original and the stego image in the case of $N = 500$ 
are shown in Fig.\ref{fig:source} 
and Fig.\ref{fig:stego}, respectively. 

Next, we compare the case of $(16, 16, 256)$-complete complementary codes 
with the cases where M-sequences and Gold sequences 
of the period 4,095 are employed. 
When M-sequences or Gold sequences are employed, 
secret messages are embedded by the conventional direct-spreading (DS) 
modulation scheme. 

\begin{figure}[!t]
  \begin{center}
    \includegraphics[scale=0.65]{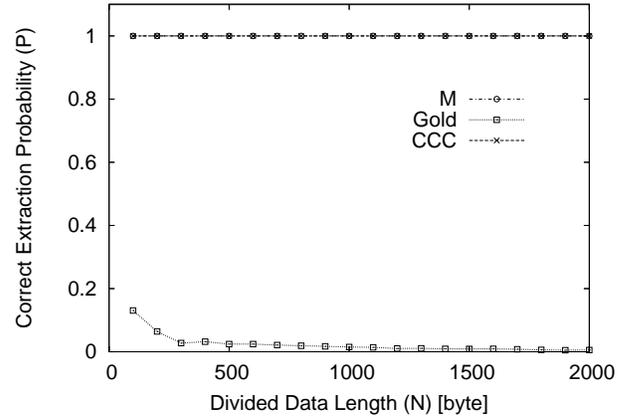}
  \end{center}
 \caption{The relation between the divided data length $N$ 
	and the correct extraction probability $P$.}
 \label{fig:result1}
\end{figure}

Figure \ref{fig:result1} shows the relation between the length of 
the divided data fragment $N$ and the correct extraction probability 
$P$, which is defined as 
\begin{equation}
  P = \frac{\rm number ~ of ~ correctly ~ extracted ~ characters}
	{\rm number ~ of ~ all ~ characters ~ in ~ the ~ text ~ data}. 
\end{equation}
The number of secret channels is fixed as $M = 1$, 
and the divided data length is varied from $N = 100$ to 2,000 bytes. 
Each curve is evaluated by 100 independent trials. 
From Fig.\ref{fig:result1}, the embedded secret messages can be 
extracted perfectly in both cases of M-sequences 
and complete complementary codes (CCC). 
On the other hand, 
the messages cannot be extracted in most cases 
when Gold sequences are employed. 
This implies that the auto-correlation properties of the employed 
sequences play important roles in extracting the embedded 
secret messages. 

\begin{figure}[tb]
  \begin{center}
    \includegraphics[scale=0.65]{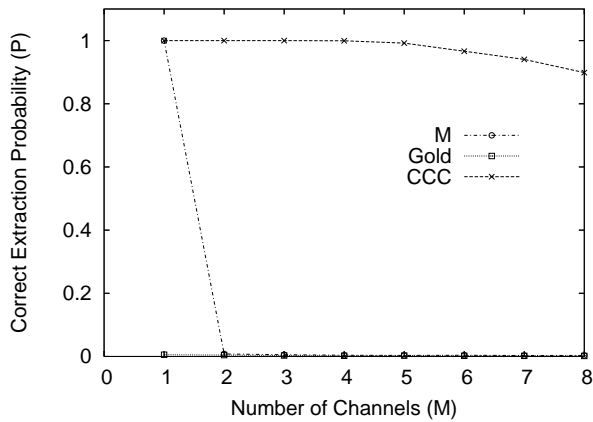}
  \end{center}
 \caption{The relation between the number of channes $M$ 
	and the correct extraction probability $P$.}
 \label{fig:result2}
\end{figure}

On the other hand, Fig.\ref{fig:result2} shows the relation between 
the number of secret channels $M$ and the correct extraction 
probability $P$ in the case where the divided data length 
is fixed as $N = 2,000$ bytes. 
The number of channels $M$ is varied from 1 to 8. 
Each curve is also evaluated by 100 independent trials. 
It is shown that the correct extraction probability $P$ decreases 
as the number of channels $M$ increases. 
However, when the plural channels are employed at the same time, 
the complete complementary code is superior to M-sequences and 
Gold sequences. 
In other words, the more embedded secret messages can be 
successfully extracted in the cases of complete complementary codes 
than those in the cases of M-sequences and Gold sequences. 
This implies that the correct extraction probability $P$ depends 
upon the cross-correlation properties of the employed sequences. 
From this result, it is expected that a large amount of secret messages 
can be embedded and extracted successfully by using the proposed 
steganography method. 

\section{Conclusions}

We proposed a steganography based on CT-CDMA communication scheme using 
complete complementary codes. 
The effectiveness of the proposed method is evaluated through the numerical 
experiments. 
It is shown that at most 600-byte text message can be embedded 
secretly into the cover image. 
In addition, when the multiple channels are used to form the stego images, 
the embedded secret messages are quite successfully extracted 
in the case of complete complementary codes than those of 
M-sequences and Gold sequences. 
It can be considered that this superiority comes from 
the good auto- and cross-correlation properties 
of complete complementary codes.  

If we embed a single secret message with large size into the cover image 
by using $M$ secret channels at the same time, it corresponds 
to the multiplexing from the viewpoints of the communication systems. 
On the other hand, if we embed $M$ different secret messages 
into $M$ mutually distinct channels, it corresponds to 
the information transmission by the multi-terminal or multi-user 
communication systems. 
In previous studies, it has been shown that complete complementary codes 
are effective for both of multiplexing and multi-user communications. 
Therefore, it can be expected that 
the proposed digital watermarking scheme and steganography method 
have the superiority to other information hiding schemes 
based upon spread spectrum techniques. 
However, the numerical results show that $PSNR$ degrades 
even for the complete complementary codes when multiple channels are 
employed at the same time. 
This could be modified if the appropriate parameters are carefully chosen 
in the embedding procedure. 

In our future study, it is quite important to investigate the effect 
of some image processing such as scaling, trimming, and JPEG compression 
or the collusion attacks, 
and propose the methods to resist against these attacks. 
It is also important to reduce the computational complexity 
in the extracting procedures when the long sequence is employed. 

\section*{Acknowledgments}

This research was partially supported by Grant-in-Aid 
for Scientific Research (C) 20293136, 2009, 
from the Japan Society for the Promotion of Science. 
One of the authors, T. Kojima sincerely thanks to 
Prof. Shinya Matsufuji of Yamaguchi University, Japan, 
and Dr. Minoru Kuribayashi of Kobe University, Japan, 
for their fruitful discussions and comments. 

\bibliographystyle{IEEEtran}
\bibliography{IEEEabrv,mybibdata_ieee}

\begin{thebibliography}{1}
\providecommand{\url}[1]{#1}
\csname url@samestyle\endcsname
\providecommand{\newblock}{\relax}
\providecommand{\bibinfo}[2]{#2}
\providecommand{\BIBentrySTDinterwordspacing}{\spaceskip=0pt\relax}
\providecommand{\BIBentryALTinterwordstretchfactor}{4}
\providecommand{\BIBentryALTinterwordspacing}{\spaceskip=\fontdimen2\font plus
\BIBentryALTinterwordstretchfactor\fontdimen3\font minus
  \fontdimen4\font\relax}
\providecommand{\BIBforeignlanguage}[2]{{%
\expandafter\ifx\csname l@#1\endcsname\relax
\typeout{** WARNING: IEEEtran.bst: No hyphenation pattern has been}%
\typeout{** loaded for the language `#1'. Using the pattern for}%
\typeout{** the default language instead.}%
\else
\language=\csname l@#1\endcsname
\fi
#2}}
\providecommand{\BIBdecl}{\relax}
\BIBdecl

\bibitem{cox97}
I.~J. Cox, J.~Kilian, F.~T. Leighton, and T.~Shamoon, ``Secure spread spectrum
  watermarking for multimedia,'' \emph{{IEEE} Trans. Image Process.}, vol.~6,
  no.~12, pp. 1673--1687, Dec. 1997.

\bibitem{hayashi2007}
N.~Hayashi, M.~Kuribayashi, and M.~Morii, ``Collusion-resistant fingerprinting
  scheme based on the cdma- technique,'' in \emph{IWSEC2007, LNCS 4752}.\hskip
  1em plus 0.5em minus 0.4em\relax Springer-Verlag, Oct. 2007, pp. 28--43.

\bibitem{aminaga2003}
H.~Aminaga, Y.~Tanada, T.~Matsumoto, and S.~Matsufuji, ``A correlation-based
  digital watermarking method using two-dimensional complementary pairs (in
  {Japanese}),'' in \emph{Tech. Rep. of IEICE, WBS2003-79}, Oct. 2003, pp.
  73--77.

\bibitem{suehiro88}
N.~Suehiro and M.~Hatori, ``{$N$-}shift cross-orthogonal sequences,''
  \emph{{IEEE} Trans. Inf. Theory}, vol.~34, no.~1, pp. 143--146, Jan. 1988.

\bibitem{suehiro00}
N.~Suehiro, N.~Kuroyanagi, T.~Imoto, and S.~Matsufuji, ``Very efficient
  frequency usage system using convolutional spread time signals based on
  complete complementary code,'' in \emph{Proc. PIMRC 2000}, London, UK, Sept.
  2000, pp. 1567--1572.

\bibitem{kojt2006}
T.~Kojima, A.~Fujiwara, K.~Yano, M.~Aono, and N.~Suehiro, ``Comparison of the
  two signal design methods in the {CDMA} systems using complete complementary
  codes,'' \emph{IEICE Trans. Fundamentals}, vol. E89-A, no.~9, pp. 2299--2306,
  Sept. 2006.

\bibitem{kojt2008}
T.~Kojima and M.~Aono, ``Properties of a convoluted-time and code division
  multiple access communication systems based upon complete complementary
  codes,'' \emph{IEICE Trans. Fundamentals}, vol. E91-A, no.~10, pp.
  2881--2884, Oct. 2008.

\bibitem{horii2009}
Y.~Horii and T.~Kojima, ``On digital watermarks based on complete complementary
  codes,'' in \emph{Proc. of IWSDA'09}.\hskip 1em plus 0.5em minus 0.4em\relax
  IEEE Press, Oct. 2009, pp. 126--129.

\bibitem{jin2009}
Y.~Jin, H.~Koga, and N.~Suehiro, ``A unified approach to construction of the
  complete complementary codes using sets of orthogonal vectors (in
  {Japanese}),'' \emph{IEICE Trans. Fundamentals}, vol. J92-A, no.~2, pp.
  105--114, Feb. 2009.

\end{thebibliography}
%
%
%

\end{document}